\documentclass[aps,twocolumn]{revtex4}
\usepackage[pdftex]{graphicx}

\usepackage{amssymb}
\usepackage{amsmath}
\usepackage{natbib}

\newcommand\br{\begin{eqnarray}}
\newcommand\er{\end{eqnarray}}
\newcommand\be{\begin{equation}}
\newcommand\ee{\end{equation}}

\newcommand\bc{\begin{center}}
\newcommand\ec{\end{center}}

\newcommand{\nn}{\nonumber \\}

\newcommand{\ket}[1]{|#1 \rangle}
\newcommand{\bra}[1]{\langle #1 |}

\renewcommand\a{\alpha}

\begin{document}
\title{Consistent gauge interaction involving dynamical coupling and anomalous current}

\author
{E. I. Guendelman \footnote{e-mail: guendel@bgu.ac.il}}
\affiliation{Physics Department, Ben Gurion University of the Negev, Beer
Sheva 84105, Israel}

\author
{R. Steiner \footnote{e-mail: roeexs@gmail.com}}
\affiliation{Physics Department, Ben Gurion University of the Negev, Beer
Sheva 84105, Israel}

\date{\today}

\begin{abstract}
We show a possible way to construct a consistent formalism where the effective electric charge can change with space and time without destroying the invariance. In the previous work \citep{GlobalSQED}\citep{GlobalSQCD} we took the gauge coupling to be of the form $g(\phi)j_\mu (A^{\mu} +\partial^{\mu}B)$ where $B$ is an auxiliary field, $ \phi $ is a scalar field and the current $j_\mu$ is the Dirac current. This term produces a constraint $ (\partial_{\mu}\phi) j^{\mu}=0 $ which can be related to M.I.T bag model by boundary condition.
In this paper we show that when we use the term $ g(\phi)j_{\mu}(A^{\mu} - \partial^{\mu}(\frac{1}{\square}\partial_{\rho}A^{\rho})) $, instead of the auxiliary field $ B $, there is a possibility to produce a theory with dynamical coupling constant, which does not produce any constraint or confinement. The coupling $ j_{\mu}^{A}(A^{\mu} - \partial^{\mu}(\frac{1}{\square}\partial_{\rho}A^{\rho})) $ where $ j_{\mu}^{A} $ is an anomalous current also discussed.
\end{abstract}

\maketitle
\section{Introduction}

There is a possibility of formulating a consistent formalism where the effective electric charge can change with space and time, such possibility have been considered in cosmological contexts.
Many papers have been published on the subject of the variation of the fine structure constant.
There are some clues that show that the structure constant has been slightly variable, although this is not generally agreed.
Bekenstein \citep{Bekenstein} has shown a different approach to formulate consistently a theory with a variable coupling constant. The Oklo natural geological fission reactor has lead to a measurement that some claim it implies the structure constant has changed by a small amount of the order of $ \frac{\dot{\alpha}}{\alpha} \approx 1\times 10^{-7} $ \citep{Uzan}.
In the work that we have done in ref \citep{GlobalSQED} and ref \citep{GlobalSQCD} we argue that dynamical Coupling Constants lead to confinement.
The dynamical Coupling Constants force us to use auxiliary field, which save the invariance of the action. This kind of auxiliary field was used in the past as for example in the work of Cornwall \citep{Cornwall} that used the term $ m^{2}(A^{\mu} +\partial^{\mu}B)^{2} $ to set an invariant mass term to a vector field. In the previous work \citep{GlobalSQED}\citep{GlobalSQCD} we took the gauge coupling to be of the form $g(\phi)j_\mu (A^{\mu} +\partial^{\mu}B)$ where $B$ is an auxiliary field and the current $j_\mu$ is the Dirac current. This term produces confinement mechanism (In section \ref{review} we review the mechanism).
In this paper we show (section \ref{no bound}) that when we use the term which Schwinger \citep{Schwinger} have found when he establish the general kinematic basis for appearance of massive vector mesons in gauge theories, instead of the auxiliary field $ B $, there is a possibility to produce a mechanism with dynamical coupling constant, which don't make any constraint or confinement. We should note that Cornwall also considered the Schwinger  projector term to replace the auxiliary field $ B $.
Next we will show the connection between the mechanism that produce confinement and that which don’t produce confinement. At the end of the article we consider quantization of our model.

\section{Review of Confining Boundary conditions in Abelian case }\label{review}
 In this chapter we review the work that we have done in ref \citep{GlobalSQED} and ref \citep{GlobalSQCD}. We review that dynamical Coupling Constants can lead to confinement.
The dynamical Coupling Constants is dynamical mostly at the boundary of the confinement and outside the boundary.  The gauge coupling has a term of the form $g(\phi)j_\mu (A^{\mu} +\partial^{\mu}B)$ where $B$ is an auxiliary field and the current $j_\mu$ is the Dirac current.
Before studying the issue of dynamical gauging, we review how the $ B $ field can be used in a gauge theory playing the role of a scalar gauge field \citep{Scalargaugefield}\footnote{for previous, but less general treatments involving scalar gauge field see \citep{Guendelman2013, Stueckelberg, Guendelman1979}}. That can be used to define a new type of convariante derivative.
Starting with a complex scalar field we now gauge the phase symmetry of $\phi$ by introducing a real, scalar $B( x _\mu)$ and two types of covariant derivatives as
\begin{equation}
\label{cov-ab}
D ^A _\mu = \partial _\mu + i e A_\mu ~~~;~~~ D ^B _\mu = \partial _\mu + i e \partial _\mu B ~.  
\end{equation}  
The gauge transformation of the complex scalar, vector gauge field and scalar gauge field have the 
following gauge transformation
\begin{equation} \label{gauge-trans}
\phi \rightarrow e^{i e \Lambda} \phi ~~~;~~~ A_\mu \rightarrow A_\mu - \partial _\mu \Lambda ~~~;~~~
B \rightarrow B - \Lambda ~.
\end{equation} 
It is easy to see that terms like $D ^A _\mu \phi$ and $D ^B _\mu \phi$, will be
covariant under Eq. \ref{gauge-trans} that is they transform the same way as the scalar field $ \phi $ and their complex conjugates will transform as $\phi^*$ does. Thus one can generate kinetic energy type terms like 
$(D ^A _\mu \phi) (D ^{A \mu} \phi)^*$, $(D ^B _\mu \phi) (D ^{B \mu} \phi)^*$, $(D ^A _\mu \phi) (D ^{B \mu} \phi)^*$,
and $(D ^B _\mu \phi) (D ^{A \mu} \phi)^*$. Unlike $A_\mu$ where one can add a gauge invariant 
kinetic term involving only $A_\mu$ (i.e. $F_{\mu \nu} F^{\mu \nu}$) this is apparently not possible to do for the
scalar gauge field $B$. However note that the term $A_\mu + \partial _\mu B$ is invariant under the
gauge field transformation alone (i.e. $A_\mu \rightarrow A_\mu + \partial _\mu \Lambda$ and 
$B \rightarrow B - \Lambda$). Thus one can add a term like $(A_\mu + \partial _\mu B)(A^\mu + \partial ^\mu B)$
to the Lagrangian which is invariant with respect to the gauge field part only of the gauge transformation
in Eq. \ref{gauge-trans}. This gauge invariant term will lead to both mass-like terms for the vector gauge
field and kinetic energy-like terms for the scalar gauge field. In total a general Lagrangian which respects 
the new gauge transformation and is a generalization of the usual gauge Lagrangian, which has 
the form
\begin{eqnarray} \label{u2}
& {\cal L} = c_1 D^A _\mu \phi (D ^{A \mu} \phi ) ^* + c_2 D^B _\mu \phi (D ^{B \mu} \phi )^* 
\nn & + c_3 D^A _\mu \phi (D ^{B \mu} \phi )^* 
 + c_4 D^B _\mu \phi (D ^{A \mu} \phi )^* - V(\phi) \nonumber \\ &
- \frac{1}{4} F_{\mu \nu} F^{\mu \nu} + c_5 (A_\mu + \partial _\mu B)(A^\mu + \partial ^\mu B)
\end{eqnarray}
where $c_i$'s are constants that should be fixed to get a physically acceptable Lagrangian where $ c_{3}=c^{*}_{4} $ and $ c_{1}\, , c_{2}\, , c_{5} $ are real.\\
At first glance one might conclude that $B(x)$ is not a physical field, it appears that one could "gauge" it away by taking $\Lambda = B(x)$ in \ref{gauge-trans}. However in the case of symmetry breaking when one introduces a complex charged scalar field that get expectation value which is not zero, one must be careful since this would imply that the gauge transformation of the field $\phi$ would be of the form $\phi \rightarrow e^{i e B} \phi$ i.e. the phase factor would be fixed by the gauge transformation of $B(x)$. In this situation one would no longer to able to use the usual unitary gauge transformation to eliminate the Goldstone boson in the case when one has spontaneous symmetry breaking. \\ Indeed in the case when there is spontaneous symmetry breaking, the physical gauge (the  generalization of the unitary gauge) is not  the gauge $B=0$, as discussed in \citep{Scalargaugefield}, it is a gauge where the scalar gauge field $B$ has to be taken proportional to the phase of the scalar field $ \theta $, with a proportionality constant that depends on the expectation value of the Higgs field according to 
\begin{equation}
\label{physical gauge}
\theta = \frac{c_5 - a e^2 \rho_0 ^2}{e \rho_0 ^2(c_1 + a)} B ~. 
\end{equation}

Also, in general there are the three degrees of freedom of a massive vector field and the Higgs field, and therefore all together five degrees of freedom.

If there is no spontaneous symmetry breaking, fixing the gauge $B=0$  does not coincide with the gauge that allows us to display that the photon has two polarizations, this gauge being Coulomb gauge. This is true even if we do not add a gauge invariant mass term (possible given the existence of the $B$ field).
By fixing the Coulomb gauge, which will make the the photon manifestly having only two polarizations, we will have already exhausted the gauge freedom and cannot in general in addition require the gauge $B=0$. So, in Coulomb gauge where the photon will have two polarizations, the 
$B$ field and in addition the two other scalars, the real and imaginary part of $\phi$ all represent true degrees of freedom, so altogether we have five degrees of freedom, the same as the case displaying spontaneous symmetry breaking. If we add a gauge invariant mass term, even when there is no spontaneous symmetry breaking (the $c_5$ term), in the gauge $B=0$ we have three polarizations of the massive vector field and still the real and imaginary parts of the complex scalar field $\phi$, still five degrees of freedom altogether.

\subsection{Confining Boundary conditions from dynamical Coupling Constants in Abelian case}
Here we argue that dynamical coupling constant can lead to confinement, so we begin with Dirac field $ \psi $ and a real scalar field $ \phi $, with the action:
\begin{eqnarray}\label{Dirac:boundary}
& S=\int{\bar{\psi}(i\gamma^{\mu}\partial_{\mu}-m+e\gamma^{\mu}A_{\mu})\psi \,d^{4}x}
\nn & - \frac{1}{4} \int{F^{\mu\nu}F_{\mu\nu}\,d^{4}x} \nn & +\int d^{4}x [ g(\phi)\bar{\psi}\gamma^{\mu}\psi(A_{\mu}+\partial_{\mu}B)\nn & +\frac{1}{2}\partial_{\mu}\phi\partial^{\mu}\phi - V(\phi)]
\end{eqnarray}
The model is  invariant under local gauge transformations
\begin{equation} \label{GTGlobalQED2}
A^\mu \rightarrow A^\mu + \partial^\mu \Lambda \textrm{;   } \ \ \ 
B \rightarrow  B - \Lambda 
\end{equation}

\begin{equation}
\psi \rightarrow exp (ie\Lambda) \psi
\end{equation} 

The Noether current conservation law for global symmetry $ \psi \rightarrow e^{i\theta}\psi $, $\theta= constant$ is,

\begin{equation}
\partial_{\mu} j^{\mu}_{N}=(\partial_{\mu})(\frac{\partial \mathcal{L}}{\partial \psi_{,\mu}}\delta\psi) = \partial_{\mu}(\bar{\psi}\gamma^{\mu}\psi)=0
\end{equation}
The  gauge field equation, containing in the right hand side the current which  is the source of the gauge field is:
\begin{eqnarray}
\partial_{\mu}F^{\mu\nu}=(e+g(\phi))\bar{\psi}\gamma^{\nu}\psi = j^{\nu}_{Source}
\end{eqnarray}
By considering the divergence of the above equation, we obtain the additional conservation law:
\begin{eqnarray}\label{bag conservation law}
& \partial_{\mu}j^{\mu}_{Source}=\partial_{\mu}(g(\phi))\bar{\psi}\gamma^{\mu}\psi + g(\phi) \partial_{\mu}(\bar{\psi}\gamma^{\mu}\psi)\nn & = \partial_{\mu}(g(\phi))\bar{\psi}\gamma^{\mu}\psi = 0
\end{eqnarray}
If we have scalar potential $ V(\phi) $ with domain wall between two false vacuum state, than because of the transition of the scalar field on the domain wall $ \partial_{\mu}(g(\phi))=\frac{\partial g(\phi)}{\partial \phi}\partial_{\mu}\phi=\frac{\partial g(\phi)}{\partial \phi}n_{\mu}f \neq 0 $.
We must conclude that $ n_{\mu}(\bar{\psi}\gamma^{\mu}\psi)\mid_{x=domain\,wall}=0 $. This means that on the domain wall there is no communication between the two sector of the domain, which give a confinement.
\section{NO CONFINING BOUNDARY CONDITIONS FROM DYNAMICAL COUPLING CONSTANTS USING
CURRENT CONSERVATION PROJECTORS}\label{no bound}
In this section we show that when we use the term which Schwinger \citep{Schwinger} had found in which he establish the general kinematic basis for appearance of massive vector mesons in gauge theories, there is a possibility to produce a mechanism with dynamical coupling constant, which doesn't make any constraint or confinement.
We now have  $ \partial_{\mu}\frac{1}{\square}(\partial_{\rho}A^{\rho}) $, instead of the auxiliary field $ B $
So the action is:
\begin{eqnarray}\label{Dirac:boundary2}
&  S=\int{\bar{\psi}(i\gamma^{\mu}\partial_{\mu}-m+e\gamma^{\mu}A_{\mu})\psi \,d^{4}x} - \frac{1}{4} \int{F^{\mu\nu}F_{\mu\nu}\,d^{4}x} 
\nn & +\int d^{4}x [ g(\phi)\bar{\psi}\gamma^{\mu}\psi(A_{\mu}-\partial_{\mu}\frac{1}{\square}(\partial_{\rho}A^{\rho}))\nn & +\frac{1}{2}\partial_{\mu}\phi\partial^{\mu}\phi - V(\phi)]
\end{eqnarray}
Where $ \square = \partial_{\mu}\partial^{\mu} $ and:
\begin{eqnarray}
\frac{1}{\square}f(x) = \int D_{F}(x - x')f(x')d^{4}x'
\end{eqnarray}
where $ D_{F}(x - x') $ is the Dirac propagator.
\\ The model is  invariant under local gauge transformations
\begin{equation} \label{GTGlobalQED3}
A^\mu \rightarrow A^\mu + \partial^\mu \Lambda 
\end{equation}
\begin{equation}
\psi \rightarrow exp (ie\Lambda) \psi
\end{equation} 

The Noether current conservation law for global symmetry $ \psi \rightarrow e^{i\theta}\psi $, $\theta= constant$ is,
\begin{equation}
\partial_{\mu} j^{\mu}_{N}=(\partial_{\mu})(\frac{\partial \mathcal{L}}{\partial \psi_{,\mu}}\delta\psi) = \partial_{\mu}(\bar{\psi}\gamma^{\mu}\psi)=0
\end{equation}
Before we going to do variation by $ A_{\mu} $ on equation Eq. \ref{Dirac:boundary2} we treat the problematic term $ \partial_{\mu}\frac{1}{\square}(\partial_{\rho}A^{\rho}) $
\begin{eqnarray}
& \int d^{4}x [
g(\phi)\bar{\psi}\gamma^{\mu}\psi(\partial_{\mu}\frac{1}{\square}(\partial_{\rho}\delta A^{\rho}))] =
\nn &\int \{g(\phi)\bar{\psi}\gamma^{\nu}\psi \,\partial_{\nu}[\int D_{F}(x-x')\partial'_{\rho}\delta A^{\rho}(x')d^{4}x']\}d^{4}x 
\nn & = \iint \{g(\phi)\bar{\psi}\gamma^{\nu}\psi \,\partial_{\nu}[ D_{F}(x-x')\partial'_{\rho}\delta A^{\rho}(x')]\}d^{4}x'd^{4}x 
\nn & =- \iint \{ \partial'_{\rho}\delta A^{\rho}(x')D_{F}(x-x')\partial_{\nu}[g(\phi)\bar{\psi}\gamma^{\nu}\psi] \,\}d^{4}x'd^{4}x
\nn & = - \int \{\partial_{\rho}\delta A^{\rho}(x)\frac{1}{\square}\partial_{\nu}[g(\phi)\bar{\psi}\gamma^{\nu}\psi]\}d^{4}x 
\nn & =  \int \{\delta A^{\rho}(x)\partial_{\rho}[\frac{1}{\square}\partial_{\nu}(g(\phi)\bar{\psi}\gamma^{\nu}\psi)]\}d^{4}x
\nn &
\end{eqnarray}

So variation on the action \ref{Dirac:boundary2} by $ A_{\mu} $ gives
\begin{eqnarray}\label{F no constreint}
& \partial_{\mu}F^{\mu\nu}=-e\bar{\psi}\gamma^{\nu}\psi
 - (\delta^{\nu}_{\mu} - \partial^{\nu}\frac{1}{\square}\partial_{\mu})(g(\phi)\bar{\psi}\gamma^{\mu}\psi)
\nn & = j^{\nu}_{Source}
\end{eqnarray}
By considering the divergence of the above equation we can see that $ \partial_{\nu}(\delta^{\nu}_{\mu} - \partial^{\nu}\frac{1}{\square}\partial_{\mu})(g(\phi)\bar{\psi}\gamma^{\mu}\psi)=0 $, so the additional conservation law is $ j^{\nu}_{Source} $:
\begin{eqnarray}\label{bag conservation law 2}
& \partial_{\nu}j^{\nu}_{Source}= 0 = e\partial_{\nu}(\bar{\psi}\gamma^{\nu}\psi)
\nn & + \partial_{\nu}(\delta^{\nu}_{\mu} - \partial^{\nu}\frac{1}{\square}\partial_{\mu})(g(\phi)\bar{\psi}\gamma^{\mu}\psi) 
\nn & = e\partial_{\nu}(\bar{\psi}\gamma^{\nu}\psi) = e\partial_{\nu}j_{N}^{\nu} = 0
\end{eqnarray}
We can see that the dynamical coupling constant didn't constraint the current.

\section{The connection between the theories that produce and do not produce confinement}
As we can see the term $ \partial_{\mu}\frac{1}{\square}(\partial_{\rho}A^{\rho}) $ is not local. This behaviour can be easily remove by considering the model in equation Eq. \ref{Dirac:boundary2} in which we constraint the auxiliary field $ B $:
\begin{eqnarray}
\square B = \partial_{\mu}A^{\mu}
\end{eqnarray} 
this constraint can easily be added to the action by Lagrange multiplier\citep{Cornwall}:
\begin{eqnarray}
- \xi (\square B - \partial_{\mu}A^{\mu})
\end{eqnarray}

There is a relate way, which also can give us the quantization of the system.
In the paper of Guendelman \citep{Guendelman1979}, he proposed an invariant electrodynamic Lagrangian and quantized it. We can generalize to the case where there are dynamical couplings so we consider the following action: :
\begin{eqnarray}\label{Dirac model}
&  S=\int{\bar{\psi}(i\gamma^{\mu}\partial_{\mu}-m+e\gamma^{\mu}A_{\mu})\psi \,d^{4}x}
\nn & - \frac{1}{4} \int{F^{\mu\nu}F_{\mu\nu}\,d^{4}x} \nn & +\int d^{4}x [ g(\phi)\bar{\psi}\gamma^{\mu}\psi(A_{\mu}+\partial_{\mu}B)\nn & +\frac{1}{2}\partial_{\mu}\phi\partial^{\mu}\phi - V(\phi)
\nn & - \partial_{\mu}A (\partial^{\mu}B + A^{\mu} )]
\end{eqnarray}
Where $ A $ is invariant under gauge transformation as $ A\longrightarrow A $.
The equation of motion (by $ A , A_{\mu} , B $) are:
\begin{eqnarray}
& \square B + \partial_{\mu}A^{\mu}=0
\end{eqnarray}
\begin{eqnarray}\label{no constreint 2}
\partial_{\mu}F^{\mu\nu} = - (e + g(\phi))\bar{\psi}\gamma^{\nu}\psi + \partial^{\nu}A 
\end{eqnarray} 
\begin{eqnarray}\label{no constreint 3}
\square A = \partial_{\mu}(g(\phi)\bar{\psi}\gamma^{\mu}\psi) = (\partial_{\mu}g(\phi))\bar{\psi}\gamma^{\mu}\psi
\end{eqnarray}

We can see that if we motivate  equation \ref{no constreint 2} and \ref{no constreint 3} we get equation with the same behaviour like equation \ref{F no constreint}:
\begin{align}
\partial_{\mu}F^{\mu\nu} = -e \bar{\psi}\gamma^{\nu}\psi - (\delta^{\nu}_{\mu} - \partial^{\nu}\frac{1}{\square}\partial_{\mu})(g(\phi)\bar{\psi}\gamma^{\mu}\psi)
\end{align}

We can see that the equation of motion of this model is the same as the equation of motion of the last model (action \ref{Dirac:boundary2}), so the models are equivalent.\\
\section{Coupling to anomalous current}
An anomalous current for example the axial vector current is defined by $ J^{5\mu} = \bar{\psi}\gamma^{\mu}\gamma^{5}\psi $ which is the classically conserved, according to Noether theorem corresponding to global part of  the gauge symmetry $ \psi \rightarrow e^{i\lambda\gamma^{5}}\psi $ on massless Dirac action. So classically for massless fermion $ \partial_{\mu}J^{\mu 5} = 0 $. Adler \citep{Adler} and Bell and Jackiw \citep{Bell} have found that the classical behaviour of the Axial vector current of massless fermion  acquires a quantum anomaly that breaks the symmetry . When calculating the second order Feynman diagrams (when we have also gauge fields) we can see that $ \partial_{\mu}J^{\mu 5} \neq 0 $. The most problematic theories that suffer from this anomaly are those where the gauge fields are coupled to the anomalous axial vector current which is not conserved. This is a problem because the breaking of the local gauge invariance produces a breaking of the non-renormalizability of the theory, and more generically an inconsistent theory.
 We are going to show that there exist a theory where the  gauge field coupled to axial vector current, which may not be conserved, but the effective current that actually couples to the gauge field is nevertheless conserved, even if it depends on the axial vector current.
 To show this, we going to take one step forward and write instead of the term $ g(\phi)\bar{\psi}\gamma^{\mu}\psi(A_{\mu} + \partial_{\mu}B) $ in equation Eq. \ref{Dirac model} a coupling to anomalous current term, namely $ g J^{5\mu}(A_{\mu}+ \partial_{\mu}B)  $ where $ g $ is the coupling constant. (In this case we omit the scalar field $ \phi $ because it does not have any relevance to this section since the source of the non-conservation is the anomaly, not the dynamical coupling constant), so the action will be:
\begin{eqnarray}\label{Dirac model2}
&  S=\int{\bar{\psi}(i\gamma^{\mu}\partial_{\mu}-m+e\gamma^{\mu}A_{\mu})\psi \,d^{4}x}
\nn & - \frac{1}{4} \int{F^{\mu\nu}F_{\mu\nu}\,d^{4}x} 
\nn & +\int  [ g J^{5\mu}(A_{\mu}+\partial_{\mu}B)- \partial_{\mu}A (\partial^{\mu}B + A^{\mu} )]d^{4}x
\end{eqnarray}
motivates the equations of motion gives:
\begin{eqnarray}\label{F no constreint2}
& \partial_{\mu}F^{\mu\nu}=-e\bar{\psi}\gamma^{\nu}\psi
 + (\delta^{\nu}_{\mu} - \partial^{\nu}\frac{1}{\square}\partial_{\mu})(gJ^{5\mu})
\nn & = j^{\nu}_{Source}
\end{eqnarray}

We can notice that the axial current is modified by  projecting out its non-conserved part, producing a source current $ j_{\nu Source} $  such that $ \partial_{nu}j_{\nu Source} = 0 $. This behaviour restores a conserved current, even when the original current may not be conserved, namely at the end $ \partial_{\mu}F_{\mu\nu} $ depends only on  a the total resulting current which is  conserved. This mechanism can also restore the conservation in a system which couples to an axial vector current which has an anomaly for example.


\section{Quantization}
We can quantize the model by following the quantization algorithm of Dirac \citep{Dirac1950,Dirac1951}.
We take a gauge invariant mass vector field Lagrangian without the interaction term:
\begin{align}
&  S=\int{\bar{\psi}(i\gamma^{\mu}\partial_{\mu}-m+e\gamma^{\mu}A_{\mu})\psi \,d^{4}x}
\nn & - \frac{1}{4} \int[{F^{\mu\nu}F_{\mu\nu}+\frac{1}{2}\partial_{\mu}\phi\partial^{\mu}\phi - V(\phi)] \,d^{4}x}
\nn & - \int d^{4}x \, [\partial_{\mu}A (\partial^{\mu}B + A^{\mu} ) - \frac{\mu^{2}}{2}(A_{\mu} + \partial_{\mu}B)^{2}]
\end{align}
The primary constraints of the system are:
\begin{eqnarray}
& \theta_{1} =  \pi^{0} = \frac{\partial L}{\partial(\partial_{0}A^{0} )} \approx 0 
\nn \\ & \theta_{2} = \partial_{i}\pi^{i} - \pi_{B} - j^{0} = 0 \\
& j^{0} = e\bar{\psi}\gamma^{0}\psi
\end{eqnarray}
and we use the gauge fixing that was propose at \citep{Guendelman1979}:
\begin{eqnarray}
& \theta_{3} =  B \approx 0 \\
& \theta_{4} = \pi_{A} + A^{0} \approx 0
\end{eqnarray}

In this case,
let define:
\begin{align}
A(x) = \int \frac{d^{3}k}{(2\pi \hbar)^{3/2}\sqrt{2k_{0}}}[a_{\vec{k}}e^{-ikx} + a^{\dagger}_{\vec{k}}e^{ikx}]
\end{align}
so
\begin{align}
& [a_{k},a^{\dagger}_{k'}]
\nn &  = -\frac{1}{2}\sqrt{k_{0}k'_{0}}\int \frac{d^{3}x'\,d^{3}x \,i}{(2\pi \hbar)^{3}}\{\frac{1}{k'_{0}}[A(x),\dot{A}(x')]e^{-i\vec{k}\vec{x}+i\vec{k'}\vec{x'}}
\nn &+ \frac{1}{k_{0}}[A(x'),\dot{A}(x)]e^{-i\vec{k}\vec{x}+i\vec{k'}\vec{x'}} \}
\nn & = - \frac{\hbar \mu^{2}}{2(2\pi\hbar)^{3}}\sqrt{k_{0}k'_{0}}[\frac{1}{k_{0}}+\frac{1}{k'_{0}}]\int d^{3}x \, e^{-i(\vec{k} - \vec{k'} )\vec{x}} 
\nn & = -\hbar \mu^{2} \delta^{3}(\vec{k}-\vec{k'})
\end{align}
We can notice that if $ \mu^{2}>0 $ implies that the state $ \a^{\dagger}_{k}\ket{0} $ has negative norm  whose existence allows the probabilities to be negative thus violating unitarity, which means that it is not physical. If $ \mu^{2}<0 $ the state $ a^{\dagger}_{k}\ket{0} $ has positive norm, but then the gauge particle (the photon) is now a tachyon, which leads to violations of causality. The only way to avoid both tachyons and negative norm particles is to have zero mass for the photon (i.e $ \mu^{2}=0 $) , which  gives an argument for the choice of a zero mass for the gauge particle. This leaves us with zero norm particles, so we have that the $ a^{\dagger}_{k}\ket{0} $ has zero norm.\\
The Feynman propagator of the scalar $ A $ field is:
\begin{align}\label{A propagator}
& \bra{0}TA(x)A(y)\ket{0}
\nn & = \frac{-\hbar}{(2\pi\hbar)^{4}} \mu^{2} \int \frac{d^{4}k}{k^{2}+i\epsilon}e^{-ik(x - y) }
\nn & = -\mu^{2} \Delta_{f}(x-y)
\end{align}
where:
\begin{align}
\Delta_{f}(x-y)= \frac{\hbar}{(2\pi\hbar)^{4}} \int \frac{d^{4}k}{k^{2}+i\epsilon}e^{-ik(x - y) }
\end{align}

if as we argue $ \mu^{2} $ is  zero then:
\begin{align}\label{eq:prop AA}
& \bra{0}TA(x)A(y)\ket{0} = 0 
\end{align}

The propagators of the interaction of the scalar field $ A $ and the vector field $ A_{\mu} $ are:

\begin{align}\label{eq:prop AAi}
& \bra{0}T\dot{A}(x)A_{i}(y) \ket{0} = \frac{i}{2}\partial_{0}\partial_{i}\Delta_{f}(x-y)
\end{align} 
and
\begin{align}\label{eq:prop AA0}
& \bra{0}TA(x)A_{0}(y) \ket{0} = \frac{i}{2}\partial_{0}\Delta_{f}(x-y)
\end{align}

The propagator of the interaction of the vector field $ A_{\mu} $ with itself is:
\begin{align}
& \bra{0}T A_{\mu}(x)A_{\nu}(y) \ket{0}
 \nn & = (g_{\mu\nu} -\frac{\partial_{\mu}\partial_{\nu}}{\mu^{2}}) \Delta_{F}(x-y,\mu^{2}) \nn & +\frac{\partial_{\mu}\partial_{\nu}}{\mu^{2}}\Delta_{F}(x-y,\mu^{2}=0) 
\end{align} 
if as we argue $ \mu^{2} $ is  zero then:
\begin{align}
& \bra{0}T A_{\mu}(x)A_{\nu}(y) \ket{0}
 \nn & = g_{\mu\nu} \Delta_{F}(x-y,\mu^{2} = 0) - \partial_{\mu}\partial_{\nu}\frac{\partial\Delta_{F}(x-y,\mu^{2}) }{\partial \mu^{2}}|_{\mu^{2} = 0}
\end{align} 
Which is:
\begin{align}
& \bra{0}T A_{\mu}(x)A_{\nu}(y) \ket{0}
 \nn & = \frac{\hbar}{(2\pi \hbar)^{4}} \int d^{4}k \, e^{-ik(x-y)} [ \frac{g_{\mu\nu}}{k^{2} + i\epsilon} - \frac{k_{\mu}k_{\nu}}{(k^{2} + i\epsilon)^{2}}]
\end{align} 

Since the case  $ \mu^{2} = 0 $, is the only case we can avoid both tachyons or negative norm states, we have found positive norm and zero norm states, which is by deﬁnition, a semi definite Hilbert space. This is a starting point to define a Hilbert space which exclude zero norm states by defining a quotient space by the kernel (which is the space of the zero norm states) which then a semi normed becomes a normed one \citep{nonstandard}. This procedure has been used for example in the Gupta Bleuler approach to quantizing QED \citep{Gupta,Bleuler}. They \citep{Gupta,Bleuler} also found that negative norm and zero norm have no physical experimental consequences.

The interaction term of our model (see equation \ref{Dirac model}) with the appropriate gauge fixing is:
\begin{align}
H_{I} = g(\phi)\bar{\psi}\gamma^{\mu}A_{\mu}\psi
\end{align}

We can write $ g(\phi) $ as Taylor series:
\begin{eqnarray}
& g(\phi) = \Sigma_{n=0}^{\infty}\frac{1}{n!}\frac{\partial^{n} g(\phi)}{\partial \phi^{n}}|_{0}\phi^{n}
\nn & = g(0) +\frac{\partial g(0)}{\partial \phi}\phi + ...
\end{eqnarray}
So the interaction diagrams of this mechanism are:
\\ For zero order of $ g $

\includegraphics{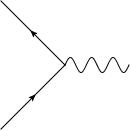}
\\ For first order of $ g $

\includegraphics{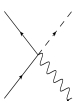}

And for second order of $ g $:

\includegraphics{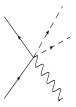}

And so on...\\
We can see that for zero order of $ g(\phi) $, the model has almost the same mechanism as QED. Because for higher order of $ g(\phi) $ the interaction involve the scalar field, then the model is not normalizable for all dimensional.
For the trivial zero order of $ g(\phi) $ the model normalizable in 4D, for first order it does not normalizable for 4D but for 3D it thus. So the question of renormalizing depends on the function $ g(\phi) $.

\section{Conclusion}
There is a possibility of formulating a consistent formalism where the effective electric charge can change with space and time, such possibility have been considered in cosmological contexts.
Many papers have been published on the subject of the variation of the fine structure constant.
There are some clues that show that the structure constant has been slightly variable, although this is not generally agreed.
Bekenstein \citep{Bekenstein} has shown a different approach to formulate consistently a theory with a variable coupling constant. The Oklo natural geological fission reactor has lead to a measurement that some claim it implies the structure constant has changed by a small amount of the order of $ \frac{\dot{\alpha}}{\alpha} \approx 1\times 10^{-7} $ \citep{Uzan}.
In the work that we have done in ref \citep{GlobalSQED} and ref \citep{GlobalSQCD} we argue that dynamical Coupling Constants lead to confinement.
The dynamical Coupling Constants force us to use auxiliary field, which save the invariance of the action. This kind of auxiliary field was used in the past as for example in the work of Cornwall \citep{Cornwall} that used the term $ m^{2}(A^{\mu} +\partial^{\mu}B)^{2} $ to set an invariant mass term to a vector field. In the previous work \citep{GlobalSQED}\citep{GlobalSQCD} we took the gauge coupling to be of the form $g(\phi)j_\mu (A^{\mu} +\partial^{\mu}B)$ where $B$ is an auxiliary field and the current $j_\mu$ is the Dirac current. This term produces confinement mechanism (In section \ref{review} we review the mechanism).
In this paper we show (section \ref{no bound}) that when we use the term which Schwinger \citep{Schwinger} have found when he establish the general kinematic basis for appearance of massive vector mesons in gauge theories, instead of the auxiliary field $ B $, there is a possibility to produce a mechanism with dynamical coupling constant, which don't make any constraint or confinement. We should note that Cornwall \citep{Cornwall} also considered the Schwinger projector term \citep{Schwinger} to replace the auxiliary field $ B $.
In this paper we have showed that when we use the term which Schwinger \citep{Schwinger} have found when he establish the general kinematic basis for appearance of massive vector mesons in gauge theories $ \partial_{\mu}\frac{1}{\square}(\partial_{\rho}A^{\rho}) $, instead of the auxiliary field $ B $, there is a possibility to produce a mechanism with dynamical coupling constant, which don't make any constraint or confinement.
As we can see the term $ \partial_{\mu}\frac{1}{\square}(\partial_{\rho}A^{\rho}) $ is not local. This behaviour can be easily remove by considering the model in equation \ref{Dirac:boundary2} and give constraint on the auxiliary field $ B $:
\begin{eqnarray}
\square B = \partial_{\mu}A^{\mu}
\end{eqnarray} 
this constraint can easily add to the action by Lagrange multiplier\citep{Cornwall}:
\begin{eqnarray}
- \xi (\square B - \partial_{\mu}A^{\mu})
\end{eqnarray}

Furthermore, We can take one step forward and and write instead of the term $ g(\phi)\bar{\psi}\gamma^{\mu}\psi(A_{\mu} + \partial_{\mu}B) $ in equation \ref{Dirac model} a coupling to anomalous current term namely $ g J^{5\mu}(A_{\mu}+ \partial_{\mu}B)  $
which give that $ \partial_{\nu}j^{\nu}_{Source} = 0 $. This behaviour restores a conserved current, even when the original current may not be conserved, namely even if $ \partial_{\mu}F^{\mu\nu} $ depends on a non conserved quantity which means that the current depends on non conserved term, still the total current are conserved. In this paper we show how to quantiszed such a system, and we calculate the propagator of each of the dynamical fields. 

\section*{Acknowledgement}
We are pleased to thank Prod. David .A Owen for his helpful reviewing of the paper and his smart comments.
We are pleased to thank Prof. Jacob Bekenstein and Prof. Yakov Itin and all the members of the GIF 4 Workshop at the Jerusalem College of Technology in Jerusalem, Israel. We also wants to thank Jerusalem College of Technology - Lev Academic Center for worm hospitality,


\bibliography{SQEDSCHWINGER}

\begin{thebibliography}{17}
\expandafter\ifx\csname natexlab\endcsname\relax\def\natexlab#1{#1}\fi
\expandafter\ifx\csname bibnamefont\endcsname\relax
  \def\bibnamefont#1{#1}\fi
\expandafter\ifx\csname bibfnamefont\endcsname\relax
  \def\bibfnamefont#1{#1}\fi
\expandafter\ifx\csname citenamefont\endcsname\relax
  \def\citenamefont#1{#1}\fi
\expandafter\ifx\csname url\endcsname\relax
  \def\url#1{\texttt{#1}}\fi
\expandafter\ifx\csname urlprefix\endcsname\relax\def\urlprefix{URL }\fi
\providecommand{\bibinfo}[2]{#2}
\providecommand{\eprint}[2][]{\url{#2}}

\bibitem[{\citenamefont{Guendelman and Steiner}(2013)}]{GlobalSQED}
\bibinfo{author}{\bibfnamefont{E.}~\bibnamefont{Guendelman}} \bibnamefont{and}
  \bibinfo{author}{\bibfnamefont{R.}~\bibnamefont{Steiner}},
  \bibinfo{journal}{arXiv}  (\bibinfo{year}{2013}), \eprint{arXiv 1311.2536}.

\bibitem[{\citenamefont{Steiner and Guendelman}(2014)}]{GlobalSQCD}
\bibinfo{author}{\bibfnamefont{R.}~\bibnamefont{Steiner}} \bibnamefont{and}
  \bibinfo{author}{\bibfnamefont{E.~I.} \bibnamefont{Guendelman}},
  \bibinfo{journal}{arXiv}  (\bibinfo{year}{2014}), \eprint{arXiv 1406.1316}.

\bibitem[{\citenamefont{Bekenstein}(1982)}]{Bekenstein}
\bibinfo{author}{\bibfnamefont{J.}~\bibnamefont{Bekenstein}},
  \bibinfo{journal}{Phys.Rev.} \textbf{\bibinfo{volume}{D25}},
  \bibinfo{pages}{1527} (\bibinfo{year}{1982}).

\bibitem[{\citenamefont{Uzan}(2003)}]{Uzan}
\bibinfo{author}{\bibfnamefont{J.}~\bibnamefont{Uzan}},
  \bibinfo{journal}{Rev.Mod.Phys.} \textbf{\bibinfo{volume}{75}},
  \bibinfo{pages}{403} (\bibinfo{year}{2003}), \eprint{hep-ph/0205340}.

\bibitem[{\citenamefont{Cornwall}(1974)}]{Cornwall}
\bibinfo{author}{\bibfnamefont{J.}~\bibnamefont{Cornwall}},
  \bibinfo{journal}{Phys.Rev.} \textbf{\bibinfo{volume}{D10}},
  \bibinfo{pages}{500} (\bibinfo{year}{1974}).

\bibitem[{\citenamefont{Schwinger}(1962)}]{Schwinger}
\bibinfo{author}{\bibfnamefont{J.~S.} \bibnamefont{Schwinger}},
  \bibinfo{journal}{Phys.Rev.} \textbf{\bibinfo{volume}{128}},
  \bibinfo{pages}{2425} (\bibinfo{year}{1962}).

\bibitem[{\citenamefont{Guendelman and Singleton}(2014)}]{Scalargaugefield}
\bibinfo{author}{\bibfnamefont{E.~I.} \bibnamefont{Guendelman}}
  \bibnamefont{and}
  \bibinfo{author}{\bibfnamefont{D.}~\bibnamefont{Singleton}},
  \bibinfo{journal}{arXiv}  (\bibinfo{year}{2014}), \eprint{arXiv 1402.7334}.

\bibitem[{\citenamefont{Guendelman}(1979)}]{Guendelman1979}
\bibinfo{author}{\bibfnamefont{E.}~\bibnamefont{Guendelman}},
  \bibinfo{journal}{Phys.Rev.Lett.} \textbf{\bibinfo{volume}{43}},
  \bibinfo{pages}{543} (\bibinfo{year}{1979}).

\bibitem[{\citenamefont{Adler}(1969)}]{Adler}
\bibinfo{author}{\bibfnamefont{S.~L.} \bibnamefont{Adler}},
  \bibinfo{journal}{Phys.Rev.} \textbf{\bibinfo{volume}{177}},
  \bibinfo{pages}{2426} (\bibinfo{year}{1969}).

\bibitem[{\citenamefont{Bell and Jackiw}(1969)}]{Bell}
\bibinfo{author}{\bibfnamefont{J.}~\bibnamefont{Bell}} \bibnamefont{and}
  \bibinfo{author}{\bibfnamefont{R.}~\bibnamefont{Jackiw}},
  \bibinfo{journal}{Nuovo Cim.} \textbf{\bibinfo{volume}{A60}},
  \bibinfo{pages}{47} (\bibinfo{year}{1969}).

\bibitem[{\citenamefont{Dirac}(1950)}]{Dirac1950}
\bibinfo{author}{\bibfnamefont{P.~A.} \bibnamefont{Dirac}},
  \bibinfo{journal}{Can.J.Math.} \textbf{\bibinfo{volume}{2}},
  \bibinfo{pages}{129} (\bibinfo{year}{1950}).

\bibitem[{\citenamefont{Dirac}(1951)}]{Dirac1951}
\bibinfo{author}{\bibfnamefont{P.}~\bibnamefont{Dirac}},
  \bibinfo{journal}{Can.J.Math.} \textbf{\bibinfo{volume}{3}},
  \bibinfo{pages}{1} (\bibinfo{year}{1951}).

\bibitem[{\citenamefont{Ng}(2010)}]{nonstandard}
\bibinfo{author}{\bibfnamefont{S.}~\bibnamefont{Ng}},
  \emph{\bibinfo{title}{Nonstandard Methods in Functional Analysis: Lectures
  and Notes}} (\bibinfo{publisher}{World Scientific}, \bibinfo{year}{2010}),
  ISBN \bibinfo{isbn}{9789814287555}.

\bibitem[{\citenamefont{Gupta}(1950)}]{Gupta}
\bibinfo{author}{\bibfnamefont{S.~N.} \bibnamefont{Gupta}},
  \bibinfo{journal}{Proc.Phys.Soc.} \textbf{\bibinfo{volume}{A63}},
  \bibinfo{pages}{681} (\bibinfo{year}{1950}).

\bibitem[{\citenamefont{Bleuler}(1950)}]{Bleuler}
\bibinfo{author}{\bibfnamefont{K.}~\bibnamefont{Bleuler}},
  \bibinfo{journal}{Helv.Phys.Acta} \textbf{\bibinfo{volume}{23}},
  \bibinfo{pages}{567} (\bibinfo{year}{1950}).

\bibitem[{\citenamefont{Guendelman}(2013)}]{Guendelman2013}
\bibinfo{author}{\bibfnamefont{E.}~\bibnamefont{Guendelman}},
  \bibinfo{journal}{Int.J.Mod.Phys.} \textbf{\bibinfo{volume}{A28}},
  \bibinfo{pages}{1350169} (\bibinfo{year}{2013}), \eprint{arXiv 1307.6913}.

\bibitem[{\citenamefont{Stueckelberg}(1957)}]{Stueckelberg}
\bibinfo{author}{\bibfnamefont{E.}~\bibnamefont{Stueckelberg}},
  \bibinfo{journal}{Helv.Phys.Acta} \textbf{\bibinfo{volume}{30}},
  \bibinfo{pages}{209} (\bibinfo{year}{1957}).

\end{thebibliography}
\bibliographystyle{apsrev}

\end{document}